\documentclass[aps,prd,twocolumn,nofootinbib]{revtex4-1}

\usepackage{graphicx}
\usepackage{color}

\newcommand{\tmtexttt}[1]{{\ttfamily{#1}}}

\begin{document}

\title{BEEC: An event generator for simulating the $B_c$ meson production at an $e^+e^-$ collider }
\author{Zhi Yang$^1$}
\author{Xing-Gang Wu$^{1}$}
\email{wuxg@cqu.edu.cn}
\author{Xian-You Wang$^{2,3}$}

\address{$^{1}$Department of Physics, Chongqing University, Chongqing 401331, P.R. China\\
$^2$ Institute of Theoretical Physics, Chinese Academy of Sciences, Beijing 100190, P. R. China\\
$^3$ National Center for Nanoscience and Technology, Beijing 100190, P.R. China}

\begin{abstract}
The $B_c$ meson is a doubly heavy quark-antiquark bound state and carries flavors explicitly, which provides a fruitful laboratory for testing potential models and understanding the weak decay mechanisms for heavy flavors. In view of the prospects in $B_c$ physics at the hadronic colliders as Tevatron and LHC, $B_c$ physics is attracting more and more attention. It has been shown that a high luminosity $e^+e^-$ collider running around the $Z^0$-peak is also helpful for studying the properties of $B_c$ meson and has its own advantages. For the purpose, we write down an event generator for simulating $B_c$ meson production through $e^+e^-$ annihilation according to relevant publications. We name it as BEEC, in which the color-singlet $S$-wave and $P$-wave $(c\bar{b})$-quarkonium states together with the color-octet $S$-wave $(c\bar{b})$-quarkonium states can be generated. BEEC can also be adopted to generate the similar charmonium and bottomnium states via the semi-exclusive channels $e^{+}+e^{-}\rightarrow |(Q\bar{Q})[n]\rangle +Q +\bar{Q}$ with $Q=b$ and $c$ respectively. To increase the simulation efficiency, we simplify the amplitude as compact as possible by using the improved trace technology. BEEC is a Fortran programme written in a PYTHIA-compatible format and is written in a modularization structure, one may apply it to various situations or experimental environments conveniently by using the GNU C compiler {\bf make}. A method to improve the efficiency of generating unweighted events within PYTHIA environment has been suggested. Moreover, BEEC will generate a standard Les Houches Event data file that contains useful information of the meson and its accompanying partons, which can be conveniently imported into PYTHIA to do further hadronization and decay simulation. \\



\end{abstract}

\maketitle

\begin{widetext}

\noindent{\bf Program summary}\\

\noindent{\it Title of program} : BEEC\\

\noindent{\it Version}:  1.0 \\

\noindent{\it Program obtained from} : CPC Program Library.\\

\noindent{\it Computer}: Any computer with Fortran compiler, the program is tested
with GNU Fortran compiler and Intel Fortran compiler.\\

\noindent{\it Operating systems} : UNIX, Linux and Windows.\\

\noindent{\it Programming language used} : FORTRAN 77/90.\\

\noindent{\it Memory required to execute with typical data} : About 2.0 MB.\\

\noindent{\it No. of bytes in distributed program, (including PYTHIA 6.4.24)} : About 1.0 MB. \\

\noindent{\it Distribution format} : Compressed tar file.\\


\noindent{\it Nature of physical problem} : Production of the charmonium, the $(c\bar{b})$-quarkonium and the bottomonium via the $e^+ e^-$ annihilation channel around the $Z^0$ peak. \\

\noindent{\it Method of solution} : The production of heavy ($Q\bar{Q'}$)-quarkonium $(Q, Q'=b, c)$ via $e^+e^-$ annihilation are estimated by using the improved trace technology. The ($Q\bar{Q'}$)-quarkonium in color-singlet 1$S$-wave state, 1$P$-wave state, and the color-octet 1$S$-wave states have been studied within the framework of non-relativistic QCD. The code with option can generate weighted and unweighted events conveniently, especially, the unweighted events are generated by using an improved hit-and-miss approach so as to improve the generating efficiency. \\

\noindent{\it Restrictions on the complexity of the problem} : The generator is aimed at the production of double heavy quarkonium through $e^+e^-$ annihilation at the $Z^0$ peak. The considered processes are those that are associated with two heavy quark jets, which could provide sizable quarkonium events around the $Z^0$ peak. \\

\noindent{\it Typical running time} : It depends on which option one choices to match PYTHIA when generating the heavy quarkonium events. Typically, for the production of the $S$-wave quarkonium states, if setting IDWTUP=$1$ (unweighted events), then it takes about 2 hour on a 2.9 GHz AMD Athlon (tm) II$\times4$ 635 Processor machine to generate $10^{5}$ events; if setting IDWTUP=$3$ (weighted events), it takes only $\sim16$ minutes to generate $10^5$ events. For the production of the $P$-wave quarkonium states, the time will be almost one hundred times longer than the case of the $S$-wave quarkonium. \\

\end{widetext}

\section{background and main idea of BEEC}

Heavy quarkonium has attracted wide attention due to its special features. It provides a good platform to study the perturbative QCD and the associated non-perturbative physics in the bound state system~\cite{hqs1,hqs2}. For example, the $B_c$ meson is a doubly heavy quark-antiquark bound state and carries flavors explicitly; it decays through weak interactions only. Thus, the $B_c$ meson can be a fruitful laboratory for testing potential models and understanding the weak decay mechanism for heavy flavors. Systematic studies of its production properties at the hadronic colliders Tevatron and LHC have been done in the literature both theoretically~\cite{bc0,bc1,bc2,bc3,tbc,tbc0,wbc1,wbc2} and experimentally~\cite{exp0,exp1,exp2,exp3,exp4,exp5,exp6,exp7}. In particular, a generator BCVEGPY for the hadronic production of the $B_c$ meson has been completed and developed in recent years~\cite{bcvegpy1,bcvegpy2}, which can be conveniently implemented into PYTHIA~\cite{pythia} for simulating $B_c$ events with high efficiency. It has been noted that at the hadronic colliders, there is much pollution from the hadronic background and many produced $B_c$ events have been cut off by the trigging condition~\cite{exp0,exp1,exp2,exp3,exp4,exp5,exp6,exp7}. Some alternative measurements would be helpful for a comprehensive study.

Comparing to the hadronic colliders, a cleaner $e^+e^-$ collider is helpful and has some advantages to perform precise measurements for certain high-energy processes. Previously, the LEP-I experiment did a try to seek the $B_c$ events, but because of its small collision energy and low luminosity, no $B_c$ events have been found there. If the incident $e^+ e^-$ collision energy is around the $Z^0$-peak and its luminosity can be reached up to ${\cal L}\propto 10^{34-36}cm^{-2}s^{-1}$ (the so-called super $Z$ factory~\cite{wjw}, or the Gigaz program suggested by the Internal Linear Collider Collaboration~\cite{ilc,gigaz}), its production rate can be raised up by several orders, then it could be observable~\cite{zbc0,zbc1,zbc2,zbc3,zbc4,eebc}. Thus, new opportunity for studying the $B_c$-meson properties at the $e^+ e^-$ colliders arises. For the sake of experimental feasibility studies, we write a generator, named as BEEC, for simulating the $B_c$ meson events at the future high luminosity $e^+ e^-$ colliders.

\begin{figure}[htb]
\includegraphics[width=0.48\textwidth]{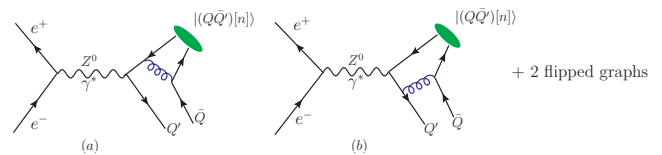}
\caption{Feynman diagrams for the heavy quarkonium production via the $e^+ e^-$ annihilation, $e^{+}+e^{-}\rightarrow |(Q\bar{Q'})[n]\rangle +Q' +\bar{Q}$, where $Q$ or $Q'$ stands for $b$ or $c$ respectively, and $[n]$ stands for the Fock states $|(Q\bar{Q'})_{\bf 1}[^1S_0]\rangle$, $|(Q\bar{Q'})_{\bf 8}[^1S_0]g\rangle$, $|(Q\bar{Q'})_{\bf 1}[^3S_1]\rangle$, $|(Q\bar{Q'})_{\bf 8}[^3S_1]g\rangle$, $|(Q\bar{Q'})_{\bf 1}[^1P_1]\rangle$ and $|(Q\bar{Q'})_{\bf 1}[^3P_J]\rangle$ (with $J=(1,2,3)$) respectively.} \label{feyfig}
\end{figure}

In the framework of the effective theory of non-relativistic QCD (NRQCD) \cite{nrqcd}, a heavy quarkonium is considered as an expansion of various Fock states. In Ref.\cite{eebc}, we have presented a detailed analysis of the $(c\bar{b})$-quarkonium production via the process $e^{+}+e^{-}\rightarrow |(c\bar{b})[n]\rangle+b+\bar{c}$, where $[n]$ stands for the dominant $(c\bar{b})$-quarkonium state within NRQCD. More explicitly, $[n]$ stands for the quarkonium states $|(Q\bar{Q'})_{\bf 1}[^1S_0]\rangle$, $|(Q\bar{Q'})_{\bf 8}[^1S_0]g\rangle$, $|(Q\bar{Q'})_{\bf 1}[^3S_1]\rangle$, $|(Q\bar{Q'})_{\bf 8}[^3S_1]g\rangle$, $|(Q\bar{Q'})_{\bf 1}[^1P_1]\rangle$ and $|(Q\bar{Q'})_{\bf 1}[^3P_J]\rangle$ ($J=(1,2,3)$) respectively. The analysis there can be extended to all the doubly heavy quarkonia $(c\bar{c})$, $(c\bar{b})$ and $(b\bar{b})$ via the process $e^{+}+e^{-}\rightarrow |(Q\bar{Q'})[n]\rangle +Q'+\bar{Q}$ ($Q$ and $Q'$ stand for $b$ or $c$), whose typical Feynman diagrams are presented in Fig.(\ref{feyfig}). The remaining two flipped diagrams are obtained by interchanging the position of the $Q'$ and the $\bar{Q}$ lines in the first two diagrams.

The leptonic part of the process can be easily dealt with, while the part for $Z^0/\gamma^* \rightarrow |(c\bar{b})[n]\rangle+b+\bar{c}$ is much more involved. Because of the emergence of massive-fermion lines in these Feynman diagrams, the analytic expression for the squared amplitude becomes too complex and lengthy under the conventional squared amplitude approach. Especially, to derive the amplitudes for the $P$-wave states, one also needs to get the derivative of the amplitudes over the relative momentum of the constitute quarks. One important way to solve it is to deal with the process directly at the amplitude level, i.e. after generating proper phase-space points, one first calculate the numerical value for the amplitudes, and then sum these values algebraically and square it to get the squared amplitude, $|{\cal M}|^2=|\sum_{i}{\cal M}_{i}|^2$; through such way, numerical simulation efficiency can be greatly improved in comparison to the usual squared amplitude technology. Moreover, under the approach, many simplifications can be done at the amplitude level due to the fermion-line symmetries and the specific properties of each heavy-quarkonium Fock states.

The improved trace technology is designed for such purpose. As an explanation, we first arrange any one of the amplitude $M_{ss^{\prime}}$ into four orthogonal sub-amplitudes $M_{i}$ according to the spins of the outgoing quark $Q$ with spin $s$ and the outgoing anti-quark with spin $s'$, then transform these sub-amplitudes into a trace form by properly dealing with the massive spinors with the help of an arbitrary light-like momentum $k_0$ and an arbitrary space-like momentum $k_1$, which satisfies $k_1^2=-1$ and $k_0\cdot k_1 =0$. The final results should be independent of $k_0$ and $k_1$, and one can choose them to be those that can maximally simply the amplitude. Then we do the trace of the Dirac $\gamma$-matrix strings at the amplitude level, which finally results in analytic series over some independent Lorentz-structures. Detailed description of the improved trace technology, together with the necessary analytical expressions, can be found in Refs.~\cite{zbc2,zbc3,eebc}. All the independent Lorentz-structures together with their coefficients are put into the generator BEEC as separate subroutines.

To do the phase-space integration, we first adopt the subroutine RAMBOS~\cite{rambos} to generate the required phase-space points, which also transforms the phase-space integration to be those variables varying within the region of $[0,1]$. In the program BEEC, a switch to choose whether using the subroutine VEGAS~\cite{vegas} is introduced. When running VEGAS, the most important samples for the squared matrix element of the process are taken first, which is in agreement with the importance sampling strategy for Monte Carlo simulations. By taking an adequate number of sampling points for the integration, the output of VEGAS could reach up to a stable result with requested statistical error.

We provide several ways in BEEC to generate the weighted and unweighted events. For theoretical studies on the heavy quarkonium production, e.g. to derive the total cross-section or various differential distributions, one can directly use the fastest way (preferred): to generate the weighted events that are distributed according to the importance sampling function generated by VEGAS running. By using BEEC in this way, some interesting properties for the $B_c$ meson production have been found~\cite{eebc}. While, for the events simulation in detector conditions, it is necessary to get the unweighted events, which are distributed according to the matrix element squared~\cite{mchep,unweig}. For the purpose, one can generate the unweighted events by directly using the PYTHIA's inner mechanism, the so-called hit-and-miss approach (von Neumann algorithm), to reject those unsatisfied events and output the allowed events. But, as is well-known, the original hit-and-miss approach is really time-consuming. Some alterations must be made to improve its efficiency.

It has been observed that the weighted events could be mapped to realistic independent unweighted events~\cite{mchep}. Thus, instead of using the time-consuming PYTHIA inner mechanism, we generate the unweighted events by using a more effective hit-and-miss technique~\cite{gxicc21}. Compared to the original method of PYTHIA, such new hit-and-miss procedure is taken in each cell of the adaptive mesh found by VEGAS. Its idea lies in that: In order to save the amount of storage space and the efficiency, we adopt the method MINT~\cite{mint} developed by the authors of POWHEG program~\cite{powheg1,powheg2}. By using the VEGAS algorithm, the MINT program performs the integration in using the SPRING-BASES subroutines and generates events with a probability proportional to the integrand in using the SPRING subroutines~\cite{basespring}. After each iteration of VEGAS running, the maximum value of the function will be stored in a file for each cell of the adaptive mesh. The multidimensional stepwise function that equals to the upper bound of the function to be integrated in each cell in fact provides an upper bound for the whole function. So, the program is to find the upper bound grid for those cells. Next, by using again the hit-and-miss technique in each cell, one can generate the points according to the original distribution. Following these procedures, we can use BEEC to generate the unweighted heavy quarkonium events effectively, and then we store the information of the unweighted event in a standard Les Houches Event (LHE) file~\cite{lhe} so as to do the further simulation.

The paper is organized as follows. In Sec.II, we show the dominant features of the generator BEEC, in which, we present its structure, its flow chart and its usage in detail. The final section is reserved for a summary.

\section{The generator BEEC}

As described in the above section, the BEEC is designed to be a specific generator for simulating the production of doubly heavy quarkonium $(c\bar{c})$, $(c\bar{b})$ and $(b\bar{b})$ at a $e^+ e^-$ collider. Since BEEC is implemented into PYTHIA as an external process, all PYTHIA subroutines can be applied conveniently. That is, one can use the PYTHIA subroutines to read the generated useful information of the heavy quarkonium and its accompanying partons, and do further hadronization and decay simulation.

According to the NRQCD framework~\cite{nrqcd}, the cross-section of the quarkonium production process, $e^{+}(p_2) + e^{-}(p_1) \rightarrow |(Q\bar{Q'})[n]\rangle(q_3) +Q^{\prime}(q_2) +\bar{Q}(q_1)$ with $Q$ and $Q'$ stand for $b$ or $c$, can be written in the following form:
\begin{equation}
d\sigma = \sum_{n} d\hat\sigma (e^{+}+e^{-}\rightarrow (Q\bar{Q'})[n] +Q' +\bar{Q}) \langle{\cal O}^H(n)\rangle,
\end{equation}
where the non-perturbative matrix element $\langle{\cal O}^H(n)\rangle$ is proportional to the inclusive transition probability of the perturbative state $(Q\bar{Q'})[n]$ into the bound states $|(Q\bar{Q'})[n]\rangle$. The short-distance differential cross section $d\hat\sigma(e^{+}+e^{-}\rightarrow (Q\bar{Q'})[n]+ Q' +\bar{Q})$ stands for the short-distance cross-section; i.e.
\begin{widetext}
\begin{eqnarray}
&&d\hat\sigma(e^{+}+e^{-}\rightarrow (Q\bar{Q'})[n]+Q'+\bar{Q}) = \frac{1}{4\sqrt{(p_1\cdot p_2)^2-m_1^2 m_2^2}} \overline{\sum}  |{\cal M}|^{2} d\Phi_3,
\end{eqnarray}
\end{widetext}
where $\overline{\sum}$ means we need to average over the spin states of initial particles and to sum over the color and spin of all final particles, and the three-particle phase space
\begin{displaymath}
d{\Phi_3}=(2\pi)^4 \delta^{4}\left(p_1+p_2 - \sum_f^3 q_{f}\right)\prod_{f=1}^3
\frac{d^3{q_f}}{(2\pi)^3 2q_f^0}.
\end{displaymath}

To obtain the total cross sections and the differential cross sections of the processes, we shall first generate the phase-space points by the subroutine RAMBOS~\cite{rambos}, which also transforms the generated four-momentum of the final particles to proper subroutines to calculate the amplitudes ${\cal M}$, and hence the squared amplitude $|{\cal M}|^{2}$, the final integration can be carried out by using the VEGAS program~\cite{vegas}.

In the following subsections, we will sequentially present the BEEC's schematic structure, flowcharts and usage in detail.

\subsection{Structure of BEEC}

BEEC is a Fortran programme written in a PYTHIA-compatible format and is written in a modularization structure, one may apply it to various situations or experimental environments conveniently. BEEC will generate a standard LHE data file that contains useful information of the meson and its accompanying partons, which can be conveniently imported into PYTHIA to do further hadronization and decay simulation.

\begin{figure}[htb]
\includegraphics[width=0.48\textwidth]{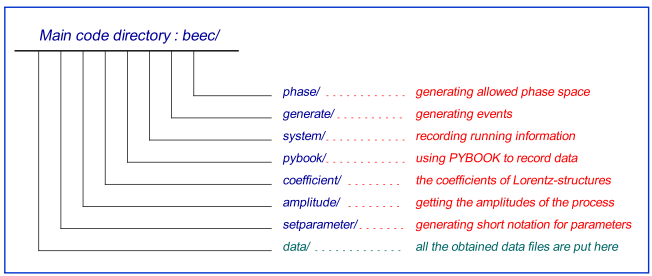}
\caption{The schematic structure for the generator BEEC.}
\label{struct}
\end{figure}

The schematic structure for the generator BEEC are shown in FIG.(\ref{struct}). All Fortran codes are organized in the main directory named as {\it beec}. In general, the generator is systematically constructed in seven modules according to their purpose. Each module contains necessary files to fulfill the specific tasks for generating events. Besides, there are three Fortran source files: parameter.F, run.F and beec.F in the main directory {\it beec}.

\begin{itemize}
\item The module {\bf generate}: It is the key module, which contains six source files: evntinit.F, genevnt.F, pythia-6.4.24.F, totfun.F, initmixgrade.F and bcpythia.F. Its main function is to initialize all input parameters for event simulation; to establish connection between BEEC and PYTHIA; to calculate the kernel for the phase-space integration with the help of the {\bf coefficient} module and the {\bf amplitude} module and to do the phase-space integration with the help of the {\bf phase} module. The file initmixgrade.F is used to initialize the importance sampling function for Monte Carlo simulation. Once the importance sampling function has been obtained by VEGAS, it can be conveniently used by initmixgrade.F for later usage without running VEGAS again. By setting proper values for the two parameters IMIX and IMIXTYPE, the required mixed quarkonium events could be generated.

\item The module {\bf phase}: it contains three source files: phase$\_$gen.F, phase$\_$point.F and vegas.F. Its purpose is to generate the allowable phase-space points and to record the importance sampling function produced by VEGAS~\cite{vegas} into a grade file (with suffix .grid) in the {\it data} subdirectory. The phase$\_$gen.F contains a reformation of the previous RAMBOS program~\cite{rambos} and can transform all the generated four-momentum of the final particles to the {\bf generate} module.

\item The module {\bf pybook}: it contains five source files: pybookinit.F, uphistrange.F, uppydump.F, uppyfact.F and uppyfill.F. Its purpose is to initialize the PYTHIA subroutine PYBOOK to record useful information of the generated events. The users may conveniently switch off this module in the main program to use his/her own ways to record the data.

\item The module {\bf setparameter}: it contains two source files: simparameter.F and uperror.F. The simparameter.F is used to simplify/optimize the input parameters that have been set in parameter.F. When the input parameters are out of allowed ranges, some typical error messages stored in uperror.F will appear on screen and the program will stop running.

\item The module {\bf system}: it contains six source files: upopenfile.F, uplogo.F, vegaslogo.F, updatafile.F, upclosegradefile.F and upclosepyfile.F. Its purpose is to open or close the record files and to print out certain running messages at the intermediate steps, which reminds the users at what step the program is running.

\item The module {\bf coefficient}: it contains four source files: coef1s0.F, coef3s1.F, coef1p1.F and coef3pj.F. It has been shown~\cite{zbc2,zbc3} that there are three independent Lorentz structures for spin-singlet color-singlet and color-octet states $|(Q\bar{Q'})_{\bf 1,8}[^1S_0]\rangle$; twelve independent Lorentz structures for spin-triplet color-singlet and color-octet states $|(Q\bar{Q'})_{\bf 1,8}[^3S_1]\rangle$; twelve independent Lorentz structures for spin-singlet $P$-wave state $|(Q\bar{Q'})_{\bf 1}[^1P_1]\rangle$; and thirty-four independent Lorentz structures for spin-triplet $P$-wave state $|(Q\bar{Q'})_{\bf 1}[^3P_J]\rangle$ (with $J=(1,2,3)$). These files store all the non-zero coefficients of all the independent Lorentz-structures for the corresponding quarkonium states.

\item The module {\bf amplitude}: it contains four files: ampy.F, ampz0.F, common.F, sqamp.F. Its purpose is to calculate the amplitude numerically according to the coefficients of Lorentz-structures listed in the {\bf coefficient} module. The amplitudes of the color-octet states $|(Q\bar{Q'})_{\bf 8}[^1S_0]g\rangle$ and $|(Q\bar{Q'})_{\bf 8}[^3S_1]g\rangle$ can be conveniently obtained from that of the color-singlet states $|(Q\bar{Q'})_{\bf 1}[^1S_0]\rangle$ and $|(Q\bar{Q'})_{\bf 1}[^3S_1]\rangle$ through proper changing of the color factors and the non-perturbative matrix elements.
\end{itemize}

Each module is equipped with a {\bf makefile} that is applied to make a library with the same name, e.g. the {\bf makefile} in the subdirectory {\bf generate} will be used by the GNU C command {\bf make} to generate {\it generate.a} located in the main directory. Once the source file has been compiled, one does not need to recompile it unless some changes have been made. A master {\bf makefile} in the main directory orchestrates all those sub-makefiles. Libraries required for the main program are listed in the {\bf LIBS} variable of the master {\bf makefile} and built automatically by invoking the sub-makefiles: \\

\begin{widetext}
\centering
\vspace{2mm}
LIBS={\it amplitude.a coefficient.a generate.a phase.a pybook.a setparameter.a system.a}
\vspace{2mm}
\end{widetext}

Under the way based on {\bf makefile}, the generator BEEC acquires good modularity and re-usability. The user can easily reform the generator to suit the needs of different experimental environment. After each running, all files generated for recording the information are put into the subdirectory {\bf data}, in which three subdirectories are for the mentioned production channels of $(c\bar{c})$, $(c\bar{b})$ and $(b\bar{b})$ respectively: all grade files for the importance sampling function are ended with the suffix `.grid'; all the files that record the used parameters and the VEGAS running information are ended with the suffix `.cs'; all the files that record the differential distributions, e.g. the transverse momentum and rapidity distributions of the heavy quarkonium, are ended with the suffix `.dat'.

\subsection{Flow charts of BEEC}

\begin{figure}[htb]
\centering
\includegraphics[width=0.35\textwidth]{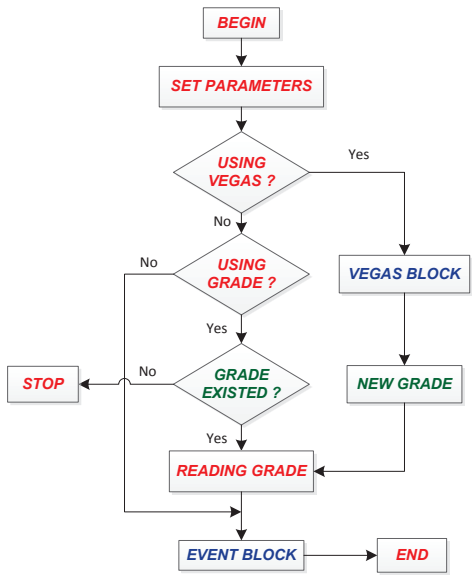}
\caption{The overall schematic flow chart of the generator BEEC.}
\label{chart1}
\end{figure}

\begin{figure*}
\centering
\includegraphics[width=0.70\textwidth]{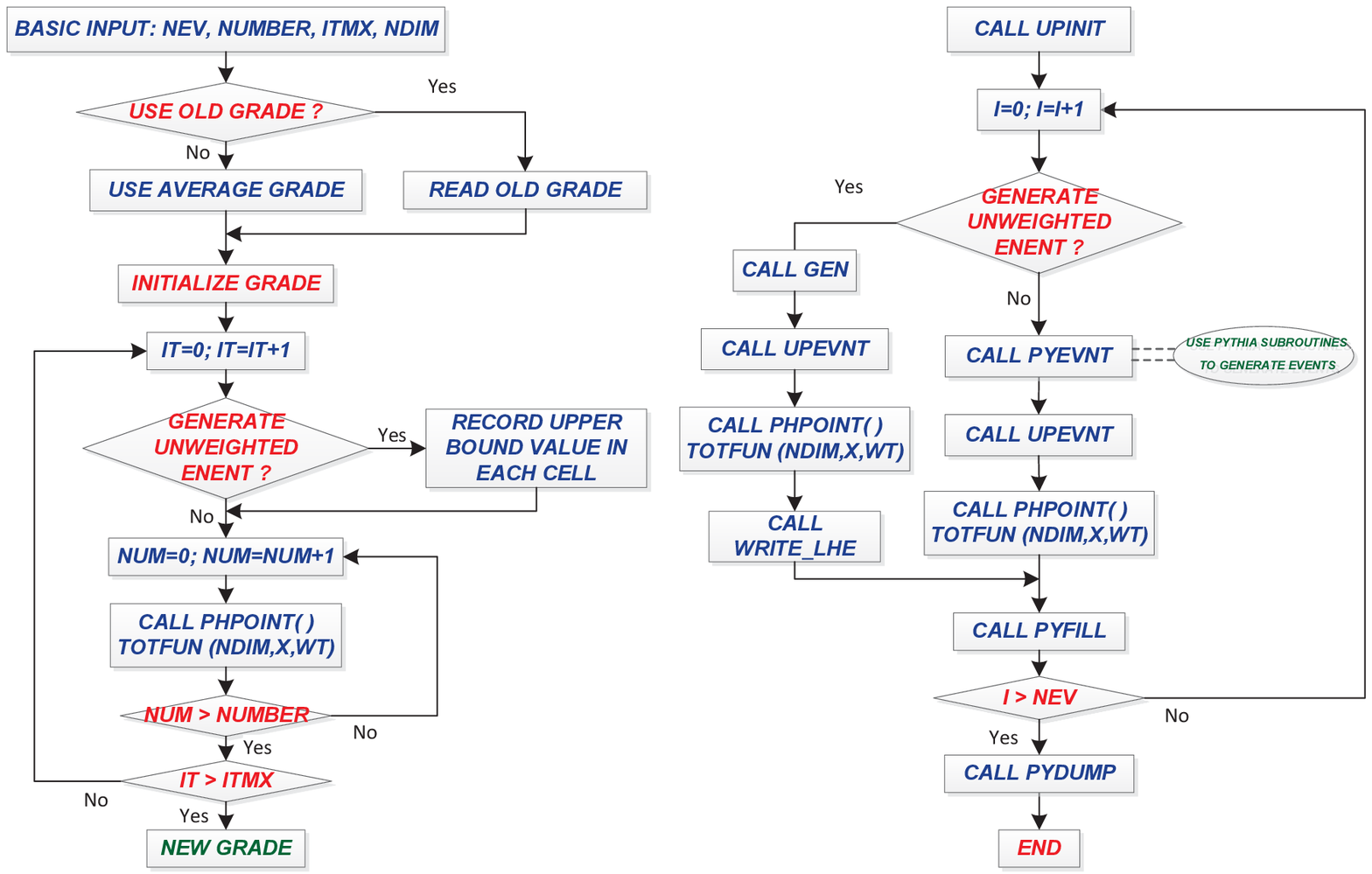}
\caption{The schematic flow chart for the {\it vegas} block (left) and
the {\it event} block (right) of the generator BEEC.}
\label{chart2}
\end{figure*}

We present the overall flow chart of the generator BEEC in Fig.(\ref{chart1}). The BEEC is dominated by two blocks, i.e. the {\it vegas} block (in module {\bf phase}) and the {\it event} block (in module {\bf generate}). The {\it vegas} block is to generate the importance sampling function. The {\it event} block is to generate events by using PYTHIA, in the way that the mentioned processes are implemented into PYTHIA as its external processes. This is achieved by properly setting the two PYTHIA subroutines UPINIT and UPEVNT. The schematic flow charts for the {\it vegas} block and the {\it event} block of the generator BEEC are show in Fig.(\ref{chart2}). Here we modify the {\it vegas} block not only to generate the sampling importance function but also to store an upper bound of the value of the cross section in each cell. The importance smapling function is used to increase the simulation efficiency, while the upper bound value will be used to generate unweighted event if the user want to do the experimental analysis and further simulation. The upper bound value in each cell is an upper bound for the cross section and also equals a multidimensional stepwise function, according to which it is easy to generate phase space points. By using the hit and miss technique, one can generate the points according to the original distribution.

The main part of the {\it vegas} block is the VEGAS subroutine. As explained in the last subsection, we adopt the MINT algorithm but with certain alterations to do the simulation. Three new variables have been added in the original VEGAS subroutine, where \tmtexttt{XINT} is the integral value for the integrand \tmtexttt{FXN} after a \tmtexttt{NDIM}-dimensional integration, the \tmtexttt{XMAX} array records the upper bounding envelope of the integrand in all cells, \tmtexttt{IMODE} is a flag: \\

\begin{widetext}
\centering
\vspace{2mm}
\noindent \fbox{\bf subroutine vegas(FXN,NDIM,NCALL,ITMX,NPRN,XINT,XMAX,IMODE)} \vspace{2mm}
\end{widetext}

\begin{itemize}

\item FXN=: the integrand;

\item NDIM=: number of integration dimensions for the generator;

\item NCALL=: maximum total number of the times to call the integrand in each iteration set by the user;

\item ITMX=: maximum number of allowed iterations set by the user;

\item NPRN=: print out level; (see VEGAS manual) $e.g.$ NPRN=$2$, when printing out only the cross section values and errors.

\item When called with IMODE=0, VEGAS performs the integration over the integrand FXN and stores the answer in a common block.

\item XMAX stands for a (NVEGBIN,NDIM) dimensional array, where NVEGBIN denotes the bin number for each coordinate. When called with IMODE=1, VEGAS will first initiate all the elements of XMAX to be ${\rm XINT}^{1/{\rm NDIM}}$, where XINT equals to the value of \tmtexttt{vegsec} that has been derived from a previous VEGAS running with IMODE=0. During the following sampling iteration, when the calculated integral value is larger than the initial XMAX(NVEGBIN,NDIM) value in a specific cell, then the value of XMAX(NVEGBIN,NDIM) for this cell will be increased by a fixed factor $f=1+1/10\,{\rm NDIM}$. After a sufficiently large number of calls, the values of XMAX(NVEGBIN,NDIM) will be stabilized for all cells. Such a final XMAX array will be stored in the same grid file as that of the importance sampling function in order to do the final simulation.

\end{itemize}

Moreover, in doing the initialization, we will call VEGAS twice by setting IMODE=0 and IMODE=1 accordingly to generate the upper bound grid XMAX and also a more precise importance sampling function. One can generate events by calling the UPEVNT subroutine according to the probability proportional to the integrand. Each event produced needs several times of iteration with three steps procedure as follows: 
\begin{itemize}
\item Calculate upper bounding function by generating a set of step-wise functions, each of them is associated with a specific coordinate (dimension); 
\item Call the \tmtexttt{phase$\_$gen} subroutine to generate a random phase-space point and calculate the integral; 
\item Judge whether such point be kept or not by using the hit-and-miss approach with the help of the upper bounding function.
\end{itemize}

In VEGAS the integral together with its numerical error are related to NCALL and ITMX. To generate full events, we suggest the user to do a test running first in order to find an effective and time-saving parameters for VEGAS. In the BEEC, we take the default values of NVEGBIN, NVEGCALL and NVEGITMX to be 300, 100000 and 10. As a subtle point, if one want to improve the precision of the generated importance sampling function, the maximum iteration number, the number of calls in each time of the iteration and the number of bins should be properly set.

\subsection{Use of BEEC}

One can run the program by using the command {\bf make} at the main directory of the generator, which automatically generate an executable file {\bf run} at the same directory. Users may change the parameters listed in the source files run.F and parameter.F to suit their own needs. Besides, the frequently changed parameters are put in a data file input.dat. For the generator BEEC, dominant parameters are listed in the following :

\begin{itemize}

\item PMB=: mass of the b quark (in unit GeV), default value $4.90$ GeV;

\item PMC=: mass of the c quark (in unit GeV), default value $1.50$ GeV;

\item PMZ=: mass of the $Z^0$ boson (in unit GeV), default value $91.1876$ GeV;

\item PME=: mass of electron (in unit GeV), default value $0.51\times10^{-3}$ GeV;

\item ECM=: collision energy of the high luminosity $e^{+}e^{-}$ collider (in units GeV), default value $91.1876$ GeV;

\item SINTHETA2=: squared value for the sine of the weak mixing angle, default value $\sin^2\theta_w=0.23119$;

\item FULLDECAY=: total decay width of $Z^0$ boson, default value $2.4952$ GeV;

\item IPROCESS=: determining which quarkonium to be generated, i.e. IPROCESS=1, 2 or 3 is to generate the $(c\bar{b})$-quarkonium, the $(c\bar{c})$-charmonium or the $(b\bar{b})$-bottomonium, respectively;

\item FBC=: radial wave function $R(0)$ for the $S$-wave quarkonium or the first derivative of the radial wave function $R^{\prime}(0)$ for the $P$-wave quarkonium. As default choice, we take their values from Refs.\cite{wforigin1,wforigin2};

\item IBCSTATE=: state of the heavy quarkonium to be generated. For the case of $(c\bar{b})$-quarkonium production, IBCSTATE=1$\sim$ 6, which are for the color-singlet $|(c\bar{b})_{\bf 1}[^1S_0]\rangle$, $|(c\bar{b})_{\bf 1}[^3S_1]\rangle$, $|(c\bar{b})_{\bf 1}[^1P_1]\rangle$ and $|(c\bar{b})_{\bf 1}[^3P_J]\rangle$ (with $J=(1,2,3)$) respectively, and IBCSTATE=7 and 8 are for the color-octet $|(c\bar{b})_{\bf 8}[^1S_0]g\rangle$ and $|(c\bar{b})_{\bf 8}[^3S_1]g\rangle$ respectively;

\item IVEGASOPEN=: whether to switch on/off the VEGAS subroutine. IVEGASOPEN=1 for using VEGAS; IVEGASOPEN=0 for not using VEGAS;

\item NUMOFEVENTS=: number of events to be generated;

\item IDWTUP=: determining how the event weights and the cross-sections should be interpreted (PYTHIA inner parameter). When IDWTUP=$3$, parton-level events have a unit weight at the input to PYTHIA, i.e. they are always accepted; while IDWTUP=$1$, events are then either accepted or rejected by using the PYTHIA inner hit-and-miss technique or the algorithm described in the above VEGAS block, so that fully generated events at the output have a common weight;

\item IDPP=: determining how the event weights and the cross-sections should be interpreted, the BEEC parameter. When setting IDPP=1 (or 3), it directly uses the way described by PYTHIA inner parameter IDWTUP= 1 (or 3) to generate the events. When setting IDPP=2, it is designed to generate the unweighted events similar to the case of IDWTUP=1; but other than applying the PYTHIA inner hit-and-miss technology, it will use our present new hit-and-miss technology to accept or reject the events in order to improve the efficiency.

\item IGENERATE=: whether to generate complete events with full decay information by applying the PYTHIA inner subroutines. IGENERATE=0, when the users wish the simulation to stop after the generation of the final states contain the required heavy quarkonium, which provides the most time-saving way and is useful for theoretical study. IGENERATE=1 if the users wish that complete events including the quarkonium decays are to be generated. In the latter case, we set IDWTUP=1 (IDPP=1);

\item MSTU(111)=: order of $\alpha_s$ for PYALPS running (a PYTHIA routine for calculating $\alpha_s$); e.g. MSTU(111)=0 for fixed $\alpha_s$ at the value PARU(111), which sets the constant value of $\alpha_s$; MSTU(111)=1 for leading order;

\item IGRADE=: whether to use the grade generated by previous VEGAS running when setting IVEGASOPEN=$0$; IGRADE=$1$ means to use; IGRADE=$0$ means not to use. This parameter is to save running time, once a grade file is generated from a previous VEGAS running, one does not need to regenerated it unless some input parameters have been changed.

\item IMIX=: whether to generate the mixed events for a specific heavy quarkonium via the production process assigned by IPROCESS. IMIX=0, when the users do not want to generate the mixed events; IMIX=1, when the users want to generate the mixed events. This is useful, since the higher Fock states' will

\item IMIXTYPE=: setting how many quarkonium states need to be mixed. Three types of mixing events have been programmed: IMIXTYPE=1, all the eight quarkonium states need to be mixed; IMIXTYPE=2, the mixed events for two color-singlet $S$-wave quarkonium states; IMIXTYPE=3, the mixed events for the four color-singlet $P$-wave plus two color-octet $S$-wave quarkonium states.

\end{itemize}

Three ways to do the Monte Carlo simulation are suggested in BEEC: one is the trivial Monte Carlo method without using VEGAS, and the other two ways are to use the importance sampling function derived by VEGAS running. For the two ways of using VEGAS, the first one is to use the existent grade (importance sampling function generated by a previous running and have been recorded in a .grid file), and the second one is to use the new grade generated by the current VEGAS running. For instance, if setting IVEGASOPEN=0 and IGRADE=1, it is to use the first importance sampling method to generate the quarkonium events, just by reading the existent importance sampling function. When using BEEC under proper options, one only needs to run VEGAS once unless those input parameters that are related to the grade have been changed.

For theoretical studies, e.g. to derive the heavy quarkonium production cross-section or various differential distributions, one can directly use the fastest way, e.g. setting IDPP=3 (equivalent to set the PYTHIA parameter IDWTUP $=3$) or setting IDPP=1 (equivalent to set IDWTUP $=1$ and IGENERATE=0) \footnote{In these cases, XMAXUP should be set to 0.}, to generate the quarkonium events. While, for the events simulation in detector conditions, it is necessary to get the unweighted events. In BEEC, the unweighted events are generated by setting IDPP=1 or 2.

\subsection{A test run}

\begin{figure}[h]
\centering
\includegraphics[width=0.45\textwidth]{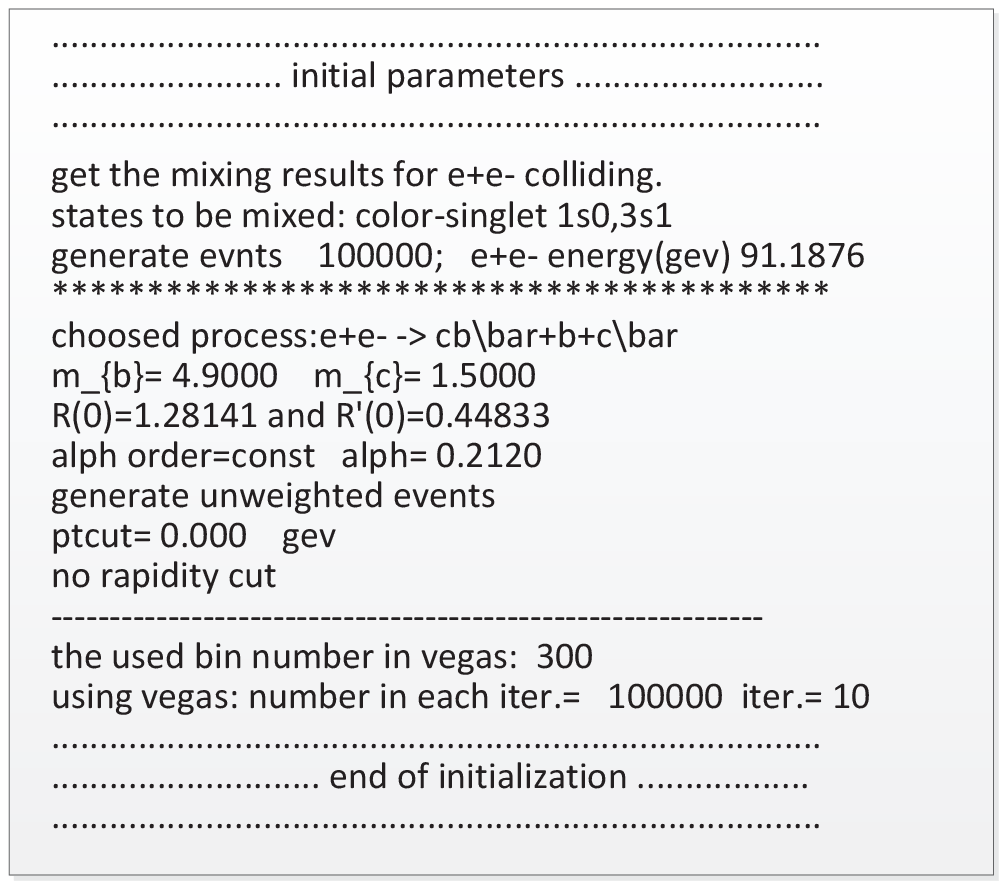}
\caption{Snapshot of the initial parameters used in the test run of the generator BEEC, which is to generate $10^5$ unweighted and mixed $(c\bar{b})$-quarkonium events for the two color-singlet S-wave states, $|(c\bar{b})_{\bf 1}[^1S_0]\rangle$ and $|(c\bar{b})_{\bf 1}[^3S_1]\rangle$.}
\label{test}
\end{figure}

\begin{widetext}
\begin{center}
\begin{figure}
\includegraphics[width=0.80\textwidth]{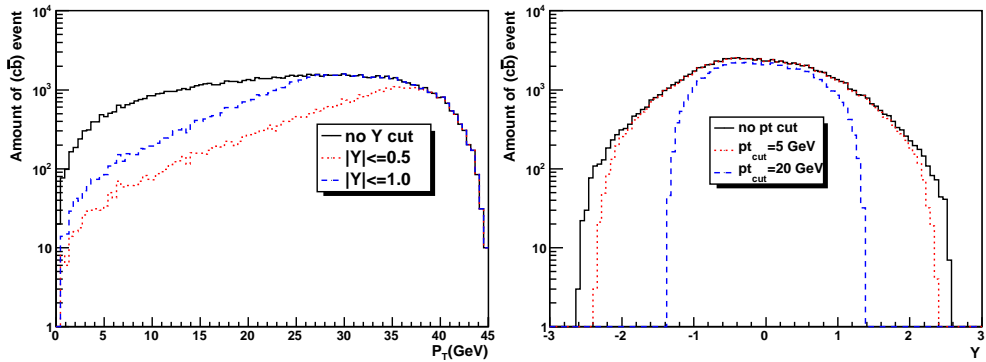}
\caption{Unweighted ($c\bar{b}$)-quarkonium events (derived by setting IDPP=2) in transverse momentum ($p_{t}$) and rapidity ($y$) distributions for the test run of the generator BEEC, which is to generate the mixed events for the states $|(c\bar{b})_{\bf 1}[^1S_0]\rangle$ and $|(c\bar{b})_{\bf 1}[^3S_1]\rangle$ (imix=1 and imixtype=2).} \label{distribution}
\end{figure}

\begin{figure}
\includegraphics[width=0.80\textwidth]{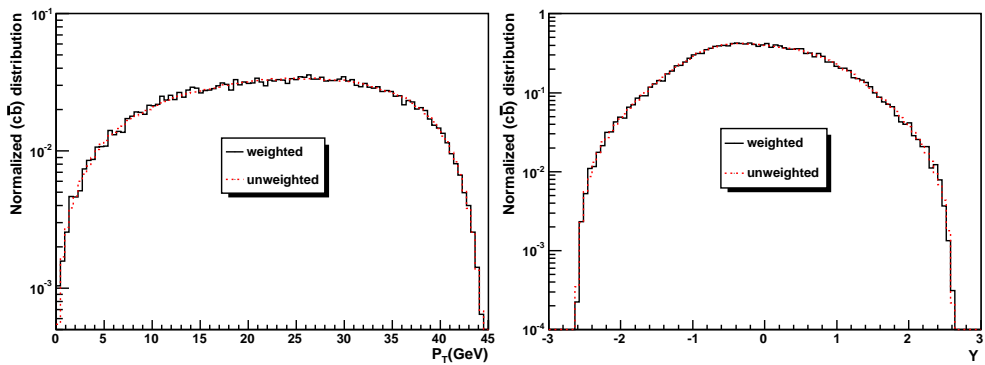}
\caption{Comparison of the $|(c\bar{b})_{\bf 1}[1^1S_0]\rangle$-quarkonium transverse momentum ($p_{t}$) and rapidity ($y$) distributions, which are derived by setting IDDP=2 (unweighted; dotted line) and IDDP=3 (weighted; solid line), respectively. } \label{compare}
\end{figure}
\end{center}
\end{widetext}

We present a test run for the production of $10^5$ mixed $(c\bar{b})$-quarkonium events in two color-singlet S-wave states, $|(c\bar{b})_{\bf 1}[^1S_0]\rangle$ and $|(c\bar{b})_{\bf 1}[^3S_1]\rangle$. It is to derive the unweighted events by using our present hit-and-miss technology by setting IDDP=2. The initial parameters for the test run are shown in Fig.(\ref{test}), which is a snapshot during the running of BEEC. We put the obtained data for the test run in the main directory as a zipped file (testdata.tar.gz), in which the running information, the total cross sections, the differential cross sections under various transverse momentum and rapidity cuts are presented. We show some typical resultant curves, e.g. $p_{t}$- and $y$-distributions with several rapidity cut and $p_{tcut}$ in Fig.(\ref{distribution}).

To be a cross-check, we also use the fastest weighted method derive the same mixed $(c\bar{b})$-quarkonium events by setting IDPP=3. For convenience, we normalize all the event curves to be the transverse momentum and rapidity distributions, which are presented in Fig.(\ref{compare}). It is found that the distributions under both the unweighted and weighted ways agree with each other. This demonstrates our present method for deriving the unweighted events is correct.

\section{Summary}

It has been found that in addition to the hadronic colliders, the super Z-factory~\cite{wjw} and the Gigaz program suggested by the Internal Linear Collider Collaboration~\cite{ilc,gigaz} will provide another good platform for studying the properties of the doubly heavy quarkoniums as the $(c\bar{b})$-quarkonium, charmonium and bottomonium. Based on our previous analysis on the production $B_c$ meson production at a high luminosity $e^+e^-$ collider~\cite{zbc2,zbc3,eebc}, we develop a generator BEEC for simulating the double heavy quarkonium events via the channel $e^{+}+e^{-}\rightarrow (Q\bar{Q'})+Q'+\bar{Q}$ (Q and $Q'$ = $c$ or $b$ respectively). BEEC is a Fortran programme written in a PYTHIA-compatible format and is written in a modularization structure, one may apply it to various situations or experimental environments conveniently. A method to improve the efficiency of generating unweighted events within PYTHIA environment has been suggested. Thus, BEEC offers a valuable tool for further experimental studies.  \\

{\bf\Large Acknowledgement:} This work was supported in part by the Fundamental Research Funds for the Central Universities under Grant No.CDJXS11100002 and No.CQDXWL-2012-Z002, by Natural Science Foundation of China under Grant No.11075225 and No.11275280, and by the Program for New Century Excellent Talents in University under Grant NO.NCET-10-0882.

\end{document}